\documentclass[aps,prl,twocolumn,amsmath,amssymb,superscriptaddress,preprintnumbers,showpacs]{revtex4-1}
\usepackage{graphicx}% Include figure files
\usepackage{dcolumn}% Align table columns on decimal point
\usepackage{bm}% bold math
\usepackage{hyperref}

\begin{document}

% Use the \preprint command to place your local institutional report
% number in the upper righthand corner of the title page in preprint mode.
% Multiple \preprint commands are allowed.
% Use the 'preprintnumbers' class option to override journal defaults
% to display numbers if necessary
%\preprint{V3rev04}

%Title of paper
\title{Wafer-scale graphene/ferroelectric hybrid devices for low voltage electronics}

\author{Yi Zheng}
   \thanks{These authors contribute equally.}
   \affiliation{Department of Physics, 2 Science Drive 3, National University of Singapore, Singapore 117542}
   \affiliation{NanoCore, 4 Engineering Drive 3, National University of Singapore, Singapore 117576}
\author{Guang-Xin Ni}
   \thanks{These authors contribute equally.}
   \affiliation{Department of Physics, 2 Science Drive 3, National University of Singapore, Singapore 117542}
\author{Sukang Bae}
     \affiliation{SKKU Advanced Institute of Nanotechnology (SAINT) and Center for Human Interface Nano Technology (HINT), Sungkyunkwan University, Suwon 440-746, Korea}
\author{Chun-Xiao Cong}
   \affiliation{School of Physical and Mathematical Sciences, Nanyang Technological University of Singapore, Singapore 637371}
\author{Orhan Kahya}
   \affiliation{Department of Physics, 2 Science Drive 3, National University of Singapore, Singapore 117542}
\author{Chee-Tat Toh}
   \affiliation{Department of Physics, 2 Science Drive 3, National University of Singapore, Singapore 117542}
\author{Hye Ri Kim}
   \affiliation{SKKU Advanced Institute of Nanotechnology (SAINT) and Center for Human Interface Nano Technology (HINT), Sungkyunkwan University, Suwon 440-746, Korea}
\author{Danho Im}
   \affiliation{Department of Chemistry, Sungkyunkwan University, Suwon 440-746, Korea}
\author{Ting Yu}
   \affiliation{School of Physical and Mathematical Sciences, Nanyang Technological University of Singapore, Singapore 637371}
\author{Jong Hyun Ahn}
   \affiliation{SKKU Advanced Institute of Nanotechnology (SAINT) and Center for Human Interface Nano Technology (HINT), Sungkyunkwan University, Suwon 440-746, Korea}
   \affiliation{School of Advanced Materials Science and Engineering, Sungkyunkwan University, Suwon 440-746, Korea}
\author{Byung Hee Hong}
   \affiliation{SKKU Advanced Institute of Nanotechnology (SAINT) and Center for Human Interface Nano Technology (HINT), Sungkyunkwan University, Suwon 440-746, Korea}
   \affiliation{Department of Chemistry, Sungkyunkwan University, Suwon 440-746, Korea}
\author{Barbaros \"{O}zyilmaz}
   \email{phyob@nus.edu.sg}
   \affiliation{Department of Physics, 2 Science Drive 3, National University of Singapore, Singapore 117542}
   \affiliation{NanoCore, 4 Engineering Drive 3, National University of Singapore, Singapore 117576}
   \affiliation{NUS Graduate School for Integrative Sciences and Engineering (NGS), Singapore 117456}

\date{\today}

\begin{abstract}
Preparing graphene and its derivatives on functional substrates may open enormous opportunities for exploring the intrinsic electronic properties and new functionalities of graphene. However, efforts in replacing SiO$_{2}$ have been greatly hampered by a very low sample yield of the exfoliation and related transferring methods. Here, we report a new route in exploring new graphene physics and functionalities by transferring large-scale chemical vapor deposition single-layer and bilayer graphene to functional substrates. Using ferroelectric Pb(Zr$_{0.3}$Ti$_{0.7}$)O$_{3}$ (PZT), we demonstrate ultra-low voltage operation of graphene field effect transistors within $\pm1$ V with maximum doping exceeding $10^{13}\,\mathrm{cm^{-2}}$ and on-off ratios larger than 10 times. After polarizing PZT, switching of graphene field effect transistors are characterized by pronounced resistance hysteresis, suitable for ultra-fast non-volatile electronics.
\end{abstract}

% insert suggested PACS numbers in braces on next line%
\pacs{72.80.Vp}
% insert suggested keywords - APS authors don't need to do this
%\keywords{}

\maketitle
As a one-atom-thick single crystal, graphene's electronic properties \cite{Geim07NatMaterReview,*Geim09ScienceReview} are closely related to its supporting substrates. SiO$_{2}$ provides excellent optical contrast, the key in discovering graphene by micromechanical exfoliation, but with critical drawbacks, such as surface roughness, high concentration of surface impurity charges, surface optical phonons, hydrophilic surface properties, and low dielectric constant ($\kappa_\mathrm{SiO_{2}}=3.9$). Such drawbacks not only limit the carrier mobility but also the dielectric gating strength by the maximum polarizability $P_{max}=\varepsilon_{0}\kappa_\mathrm{SiO_{2}} E_{max}\approx 1.7\,\mu C/cm^{2}$, where $E_{max}\approx0.5\,V/nm$ is the breakdown field of SiO$_{2}$. Substantial progresses in replacing SiO$_{2}$ have already been made, such as significant mobility enhancement of single-layer graphene on boron nitride \cite{Hong10NatureNano}, and non-volatile polymer (top) gating of single-layer graphene \cite{Zheng09APLGrapheneMemory,ZHENG10PRL_GFeFETs}. However, efforts in this direction are in general constrained by the difficulty of exfoliating and identifying in particular single and bilayer graphene on different substrates.

The rapid progresses in Copper-based chemical vapor deposition methods (Cu-CVD) have now made wafer-scale graphene synthesis and graphene transfer feasible both for single-layer graphene (SLG) \cite{Ruoff09ScienceCuCVDgraphene,Hong10NatureNano} and bilayer graphene (BLG) \cite{ZhongNanoLett_BLCuCVD2010}, providing great advantages in substrate engineering of graphene for exploring new physics and functionalities \cite{Hone10NatureNano_BN,Giovannetti07PRB_BoronNitide,Heersche07Nature_superconductivity,Zheng09APLGrapheneMemory,Novoselov09ScienceGraphane,Michetti10Nanolett_FerroProximity}. With respect to substrates, ferroelectric materials are unique both in non-volatile gating \cite{Zheng09APLGrapheneMemory} and high polarizability up to 100 $\mu\mathrm{C/cm^2}$ ($6\times10^{14}\,\mathrm{cm^{-2}}$ in charge density) \cite{Vrejoiu06SinglePZT}, 60 times larger than SiO$_{2}$. With such high gating strength, it is possible to heavily dope graphene beyond the linear band dispersion regime ($\sim1$ eV) and reach the van Hove singularities \cite{Neto09RevModernPhys}. Such high doping, which in contrast to electrolyte gating \cite{Pachoud10EPL_Electrolyte} is gate-tunable even at liquid helium temperature, may also be of great importance for verifying the recent theoretical prediction of strong electron-phonon interactions and high-temperature superconductivity in graphane and related materials \cite{SaviniPRL10_Graphane_superconductivity}. For graphene electronics, this level of gating strength may enable the opening of a sizeable \textit{non-volatile} bandgap up to $\sim300$ meV \cite{Geim07PRLBLBandgap} in bilayer graphene field effect transistors \cite{ZhangYB09NatureBandgap,*Avouris10NanoLett}. This is critical not only for achieving high current on-off ratio $>10^{4}$ for logic operations but also for improving $\Delta R/R$ for memory device applications. Equally important, it can significantly reduce the switching voltage to below 1 V while exceeding the highest doping by SiO$_{2}$ gating ($10^{13}\,\mathrm{cm^{-2}}$) \cite{Schwierz10NatureNano}.

In this letter, we demonstrate the device operation of Cu-CVD single-layer and bilayer graphene field effect transistors on ferroelectric Pb(Zr$_{0.3}$Ti$_{0.7}$)O$_{3}$ (PZT) substrates. Transistor and non-volatile memory operations have been realized by controlling PZT polarization magnitude. The ultra-high $\kappa$ of PZT in the linear dielectric regime allows graphene field effect transistors to be switched on and off within $\pm1$ V with maximum doping exceeding $10^{13}\,\mathrm{cm^{-2}}$. After polarizing PZT, the switching of graphene field effect transistors are characterized by a pronounced resistance hysteresis, ideal for ultra-fast non-volatile memory.

Large-scale graphene used in this study was synthesized by the CVD method on pure copper foils \cite{Ruoff09ScienceCuCVDgraphene,Hong10NatureNano}. By controlling the post-growth annealing time, graphene with high bilayer coverage of up to 40\%, ideal for comparing the performance of both systems, are synthesized. Subsequently, CVD graphene was transferred to 360 nm PZT, using the method introduced by Li \textit{et al} \cite{Ruoff09NanoLettTransfer,Hong10NanoLett_WaferScaleTransfer}. Standard e-beam patterning and metallization was used to fabricate $~ 3$ micron size graphene ferroelectric graphene field effect transistors (GFeFETs). The GFeFETs were then electrically characterized from room temperature (RT) to 3 K in vacuum in a four-contact configuration using lock-in amplifiers.

\begin{figure}
\begin{center}
\includegraphics[width=3.1in]{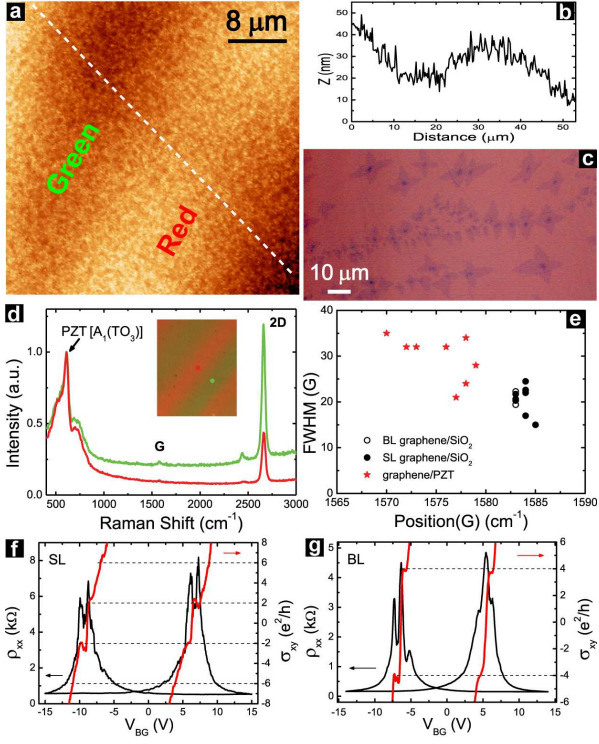}
\end{center}
\caption{(a) AFM of 360 nm PZT thin film. (b) AFM cross-section of PZT surface. (c) Optical image of high bilayer coverage graphene on SiO$_{2}$. The same batch graphene is transferred on PZT. (d) Raman spectra of Cu-CVD graphene on PZT, showing multiple reflection interference induced enhancement in $2D$ intensity. (e) FWHM and peak positions of Raman $G$ peaks of Cu-CVD graphene on PZT and SiO$_{2}$, showing significant substrate-induced strain on PZT. (f) and (g) QHE of CVD GFeFETs on PZT, showing the SLG/BLG quantization plateaux of $(N+1/2)4e^{2}/h$ and $4Ne^{2}/h$ respectively. The pronounced hysteresis in both $\rho_{xx}$ and $\sigma_{xy}$ is introduced by the ferroelectric gating.} \label{Fig01}
\end{figure}

Fig. \ref{Fig01}a shows the surface morphology of our PZT thin films measured by atomic force microscopy (AFM). PZT has periodic thickness variations of $\sim30$ nm at a typical width of $35\,\mu$m. These are easily seen as red and green stripes in optical microscopy (Inset of Fig. \ref{Fig01}d). Cu-CVD graphene transferred on PZT shows selective enhancement in Raman $2D$ intensity due to multiple reflection interference \cite{Shen08APL_Interference,*Cheong09PRB_Interference}. Raman also indicates significant substrate-induced strain in Cu-CVD graphene on PZT. As shown in Fig. \ref{Fig01}e, $G$ peaks of Cu-CVD graphene on PZT show a noticeable red shift of $\sim10\,\mathrm{cm^{-1}}$ and broadening of full width at half maximum (FWHM), compared to CVD graphene on SiO$_{2}$. Using the $G$ red shift, we estimate the PZT-induced strain to be $\sim0.2\%$ \cite{Ferrari09PRB_Strain}. This implies that Cu-CVD graphene adapts to the polycrystalline surface of PZT after transfer, which may provide a lithography free approach for substrate engineering of local strain in graphene \cite{Neto09PRLStrain}. Note that by reducing the thickness of PZT to 120 nm, SLG and BLG are both optically and Raman distinguishable. However, thin PZT films usually have much larger leakage currents. In this study, we use quantum Hall effect measurements to determine the layer number of graphene. Typical QHE for single-layer and bilayer CVD GFeFET on PZT is shown in Fig. \ref{Fig01}f and \ref{Fig01}g respectively. The characteristic quantization sequences of $(N+1/2)4e^{2}/h$ for SLG and $4Ne^{2}/h$ for BLG demonstrate the high quality of our Cu-CVD graphene.

\begin{figure}
\begin{center}
\includegraphics[width=3.2in]{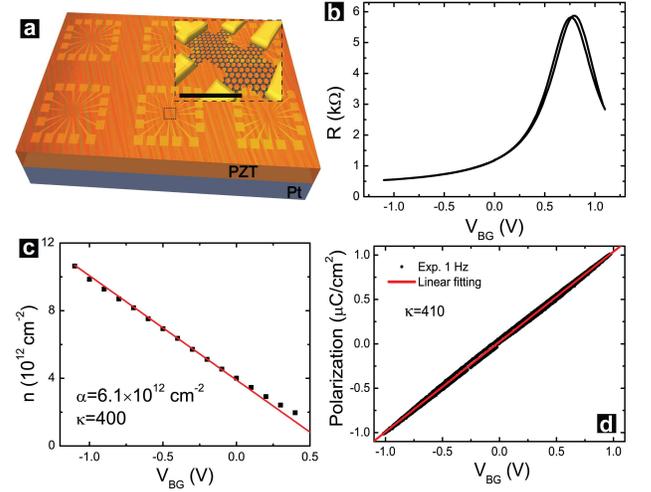}
\end{center}
\caption{(a) Cu-CVD GFeFET arrays on PZT. Inset: Schematic of an individual Hall bar device. Scale bar: 2 $\mu$m. (b) RT $R$ vs $V_\mathrm{BG}$ of a GFeFET in the linear dielectric regime of PZT. Typical mobility is $\sim 2000\,\mathrm{cm^{2}V^{-1}s^{-1}}$. (c) Linear doping vs $V_\mathrm{BG}$ relation in the linear regime with a doping coefficient of $\alpha=6.1 \times 10^{12} \mathrm{cm^{-2}V^{-1}}$ and $\kappa= 400$. (d) RT polarization measurements of PZT thin film in the linear dielectric regime using a GFeFET as the top electrode.} \label{Fig02}
\end{figure}

\begin{figure}
\begin{center}
\includegraphics[width=3.2in]{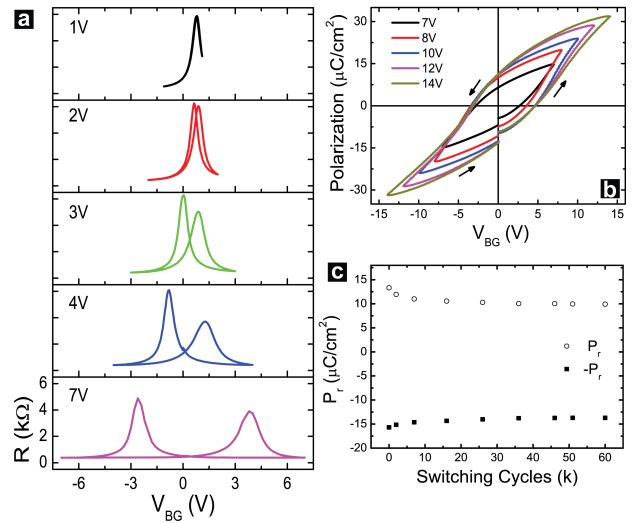}
\end{center}
\caption{(a) The evolution of $R$ vs $V_\mathrm{BG}$ hysteresis as a function of maximum $V_\mathrm{BG}$. The asymmetry in two resistance peaks are induced by the asymmetrical polarization hysteresis of PZT. (b) RT polarization measurements of PZT thin film in the polarized regime. (c) Fatigue test of PZT thin film using CVD GFeFET as the top electrodes.} \label{Fig03}
\end{figure}

In Fig. \ref{Fig02}a, we show a wafer-scale array of Cu-CVD GFeFETs on PZT. Fig. \ref{Fig02}b shows the typical resistance vs gate voltage characteristics ($R$ vs $V_\mathrm{BG}$) of GFeFETs without polarizing the PZT thin film by limiting $V_\mathrm{BG}$ below 1.1 V. In this linear dielectric regime, GFeFETs exhibit high on/off ratios exceeding 10 times with negligible $R$ vs $V_\mathrm{BG}$ hysteresis at ultra-low operating voltages previously known only from electrolyte gated samples. Hall measurements yield a linear doping vs $V_\mathrm{BG}$ relation of $n=\alpha V_\mathrm{BG}$, with $\alpha=6.1\times10^{12}\,\mathrm{cm^{-2}V^{-1}}$ (Fig. \ref{Fig02}c). This doping coefficient translates into a $\kappa$ as high as 400 using the electrical displacement continuity equation at the graphene/PZT interface \cite{Zheng09APLGrapheneMemory,ZHENG10PRL_GFeFETs}. The high doping coefficient and $\kappa$ are further confirmed by polarization measurement on the PZT thin film using the GFeFET as the top electrode (Fig. \ref{Fig03}c). Compared to the previous literature report of GFeFETs on 400 nm epitaxial Pb(Zr$_{0.2}$Ti$_{0.8}$)O$_{3}$ using multilayer graphene \cite{Zhu09PZTPRL,*Zhu10PZTAPL}, the doping coefficient in our CVD GFeFETs on PZT is almost 6 times higher. The difference in $\kappa$ and doping coefficient is most likely due to the different compositions of the PZT thin films. Indeed, by substitutional doping of Pb by Lanthanum (La) and by fine tuning the ratio between Zr and Ti, we have observed a much enhanced $\kappa$ of $\sim2000$ (not shown). Note that GFeFETs on PZT have a very broad transition area near the Dirac point, manifested by significant deviation from linear $n$ vs $V_{BG}$ relation below $3\times10^{12}\,\mathrm{cm}^{-2}$ (Fig. \ref{Fig02}b and \ref{Fig02}c). This indicates the electron-hole puddle intensity of graphene on PZT is an order-of-magnitude higher than graphene on SiO$_{2}$. The strong charge inhomogeneity in graphene on PZT is likely due to the unpolarized surface dipoles of ferroelectric thin films.

Beyond the linear regime ($V_\mathrm{BG}>1.1\,\mathrm{V}$), the polarization of PZT leads to a pronounced hysteresis in $R$ vs $V_\mathrm{BG}$ (Fig. \ref{Fig03}a). The increasing $P\mathrm{_{r}}$ not only increases the separation between the two resistance peaks, but also decreases the resistance minimum. This is because that in the polarized regime, dipole charges on ferroelectric are aligned along the same direction and flip as a single domain. Such domain flipping of dipole charges effectively mimics the clustering of organic residues, which are expected to reduce long-range scattering in CVD graphene \cite{Geim08PRB_CISClustering}. Indeed, after fully polarizing the PZT thin film, there is a factor of $\sim 2$ enhancement in mobility to $\sim 4000\,\mathrm{cm^{2}V^{-1}s^{-1}}$. The resistance hysteresis in Fig. \ref{Fig03}a can be utilized for non-volatile memory and data storage applications \cite{Zheng09APLGrapheneMemory}. Compared to the ferroelectric polymer used in Ref. \cite{Zheng09APLGrapheneMemory}, PZT allows for a significantly lower device operating voltage ($<1$ V), much faster switching speed ($<$ ns), and ultra-high endurance ($10^{10}$ cycles). In Fig. \ref{Fig03}c, we show the fatigue test ($\pm10$ V) of PZT thin films using a GFeFET as the top electrode. The nearly constant $P_\mathrm{r}$ indicates that graphene can effectively stop metal in the top layer migrating into PZT, which is the main degradation mechanism of inorganic ferroelectric. The slight degradation during the first 10k cycles is likely due to the low work function aluminum, which may contact exposed PZT surface during the wire bonding process.

In conclusion, the combination of high-quality Cu-CVD graphene and functional substrates will greatly speed up the studies of all graphene-based electronics. We demonstrate the wafer-scale patterning and device operations of Cu-CVD graphene-ferroelectric field-effect transistors on PZT substrates, integrating both transistor and non-volatile memory functionalities on the same chip by controlling the local ferroelectric polarization magnitude. In the linear regime of PZT, we demonstrate ultra-low voltage operations of GFeFETs within $\pm1$ V, which can be used as controlling transistors for addressing and reading/writing of memory unit cells. After polarizing PZT, the hysteretic switching of GFeFETs are ideal for ultra-fast non-volatile data storage. To fully utilize the switching speed of PZT, a constant doping is required to electrostatically ``biased'' the symmetrical ferroelectric doping hysteresis and create two distinct resistance states \cite{Zhu09PZTPRL}. This can be realized by non-destructive charge-transfer doping via the deposition of low work function materials on the top surface of GFeFETs \cite{Chen10APL_MoO3EG}.

\begin{acknowledgements}
This work is supported by the Singapore National Research Foundation grants NRF-RF2008-07, NRF-RF2010-07 and NRF-CRP (R-143-000-360-281), MOE2009-T2-01-037, NUS/SMF grant, US Office of Naval Research (ONR and ONR Global), and by NUS NanoCore.
\end{acknowledgements}

\end{document}